\documentstyle[12pt,epsf]{article}
\begin{document}
\textwidth 22 true cm
\textheight 25 true cm
\baselineskip=10pt
\renewcommand{\baselinestretch}{1.0}
\renewcommand{\theequation}{\arabic{equation}}
\title{{Coupling between Smectic and Twist Modes}
\\{in Polymer Intercalated Smectics}}
\author{ R.Podgornik$\dagger$ and B. \v Zek\v s $\ddagger$
\\{$\dagger$ Department of Physics, Faculty of Mathematics and Physics,}
\\{ University of Ljubljana}
\\{$\ddagger$ Institute of Biophysics, Medical Faculty, University of Ljubljana}
\\{$\dagger \ddagger$ Dept. of Theoretical Physics, ``J.Stefan'' Institute, Ljubljana} 
\\{Slovenia}
}
\begin{titlepage}
\baselineskip=8pt
\maketitle
\begin{abstract}
\small We analyse the elastic energy of an intercalated smectic where 
orientationally ordered polymers with an average orientation varying
from layer to layer are intercalated between smectic planes. The
lowest order terms in the coupling between polymer director and
smectic layer curvature are added to the smectic elastic
energy. Integration over the smectic degrees of freedom leaves an
effective polymer twist energy that has to be included into the total
polymer elastic energy leading to a fluctuational renormalization of
the intercalated polymer twist modulus. If the polymers are chiral
this in its turn leads to a renormalization of the cholesteric pitch.
\end{abstract}
\end{titlepage}

Recent elucidation of the structure of DNA - cationic lipid complexes
\cite{dnacl1,dnacl2} has brought forth quite a few unsuspected features of this
macromolecular aggregate. It appears that cationic lipids in the
complex retain their preferred packing characterised by a
multilamellar 3-D smectic order while the oppositely charged DNA gets
intercalated in between the lipid bilayer smectic planes. The
intercalated DNA appears to be packed with a 2-D smectic order of
relatively small domain sizes that are probably coupled to the order
in the neighboring intercalated DNA layers \cite{dnacl2}. The carefull
X-ray diffraction studies leave no ambiguity as to the fact that both
components of the complex - cationic lipids as well as DNA are
ordered.

The ordering tendencies giving rise to this complex aggregate
structure are due partly to the fairly well understood interactions
between lipid bilayers in aqueous solutions where van der Waals
attraction competes with electrostatic and hydration forces, augmented
by entropic repulsion forces originating in elastic fluctuations of
the lipid bilayers constrained to a multilamellar stack
\cite{adrian1}. Similarly interactions between DNA molecules in the
bulk that have only recently come under closer experimental as well as
theoretical scrutiny, appear to be dominated by repulsive forces of
electrostatic as well as hydration origin, this time too augmented
through entropic mechanism very similar to the one operating in
multilamellar lipid systems \cite{prl}. The interaction between the two
constituents of the complex are probably dominated by the
electrostatic attraction between DNA and cationic lipids possibly
modified by elastic shape fluctuations of the DNAs intercalated
between positively charged layers of lipids and possibly by the forces
mediated by the lipid bilayer elasticity due to local deformations
induced by the close proximity of intercalated DNA. More work is
certainly needed to asses the relative importance of all these
mechanisms in bringing about the stability of the DNA - cationic lipid
complex.

Compared to the phases existing in the bulk the lipid subphase does
not appear to be substantially modified. It has the same structural
geometry as the one found with other lipids in the bulk. DNA is in
this respect very much modified. At effective interhelical spacings
found in the DNA - CL complex \cite{dnacl2}, DNA in the bulk would be
either in the line hexatic phase or within the cholesteric phase
\cite{pnas}. Very little of this bulk order persists in intercalated
DNA that is forced into effectively 2-D layers intercalated between
lipid bilayers. The positional order does not change qualitatively if
we consider only DNA intercalated within a single layer. They are both
short ranged \cite{dnacl2}. The orientational order is changed more
drastically if the state of affairs in the bulk and in the DNA-CL
complex are compared. It is nevertheless the apparent total absence of
the cholesteric order in the complex that seems to us the most
baffling. Apart from very tentative statements \cite{joachim} that
cholesteric structures of extremely large pitch ( $\sim$ mm) can
sometimes be detected in the complex the chiral nature of the DNA
molecule makes no imprint on the structure of this macromolecular
aggregate.

It is our goal in this contribution to investigate the interaction
between orientational order of intercalated polymers and the smectic
degrees of freedom of intercalating lipid bilayers. We propose a
simple theory of the effect that the coupling between polymer ({\sl
i.e.} DNA) orientational ordering within the intercalated layers and
smectic order of these layers can have on the effective twist elastic
constant of the polymer layers. If in addition the intercalated
polymers are chiral, this theory for the first time introduces a
comprehensive mechanism for coupling between smectic and cholesteric
degrees of freedom leading to a fluctuational renormalization of the
cholesteric pitch of the polymer subsystem.

{\bf\sl Model} We will consider a simplified model of an intercalated
smectic phase where polymers within a {\sl single} layer are supposed
to be completely orientationaly ordered, see Fig.1. This is not
unrealistic as the domains of order in this and a similar system,
where only a single layer of DNA is adsorbed to a cationic lipid
bilayer, are quite large \cite{jie}. We will presume that the director
of the polymers ${\bf n}(n;{\mbox{\boldmath $\rho $}} )$ within an intercalated layer is a
constant. We introduced $n$ as the height index of the layer (layers
are assumed to have the average positions at $z_n = n\times d$ where
$d$ is the average layer - layer separation) while ${\mbox{\boldmath $\rho $}} = (x,y)$ is
the transverse radius vector. We have thus effectively limited
ourselves to a mean-field approximation within a single layer.

We will construct an elastic free energy of this system, assuming a
general dependence of the polymer director on the position {\sl i.e.}
height index $n$, of the layer. Since the polymer orientational order
can interact with the curvature energy of each layer we first have to
construct all the scalar invariants that can be composed with ${\bf
n}$ and the 2nd fundamental form of a single smectic layer.

If one defines the 2nd fundamental form of Gauss of the n-th layer
with local displacement described within the Monge parametrization
$(x,y, \zeta_n(x,y))$ as
\begin{equation}
K_{ik}(n;{\mbox{\boldmath $\rho $}} ) = \frac{\partial^2\zeta_n({\mbox{\boldmath $\rho $}} )}{\partial x_i\partial x_k}
\label{curv}
\end{equation}
where the indices $i,k$ can have values $i,k = 1,2$ with $x_1 = x$ and
$x_2 = y$, then the lowest order scalar invariants which can be built
from the director ${\bf n}$ and the tensor Eq.\ref{curv} of that layer
are three \cite{david}
\begin{eqnarray}
&K_{ii}K_{ik} n_i n_k& \nonumber\\
&K_{li}K_{ik} n_l n_k& \nonumber\\
&K_{ik}K_{lm} n_i n_k n_l n_m.&
\end{eqnarray}
Only two of these invariants are linearly independent if we ignore the
terms containing Gaussian curvature. The curvature elastic energy of
n-th surface, assuming that within each layer the director of the
polymers is a constant ${\bf n}(n;{\mbox{\boldmath $\rho $}}) ={\bf n}(n)$, can be written
as
\begin{eqnarray}
H_n(K_{ik};{\bf n}) &=& {\textstyle{1\over 2}} K_c \int \left( Tr
K_{ik}(n;{\mbox{\boldmath $\rho $}} )\right)^2 d^2{\mbox{\boldmath $\rho $}} + \nonumber\\ &+&
{\textstyle{1\over 2}} a \int K_{li}(n;{\mbox{\boldmath $\rho $}} )K_{ik}(n;{\mbox{\boldmath $\rho $}} )
n_l(n) n_k(n) d^2{\mbox{\boldmath $\rho $}} + \nonumber\\ &+&
{\textstyle{1\over 2}} b \int K_{ik}(n;{\mbox{\boldmath $\rho $}} )K_{lm}(n;{\mbox{\boldmath $\rho $}})
n_i(n) n_k(n) n_l(n) n_m(n) d^2{\mbox{\boldmath $\rho $}}
\nonumber\\ ~
\label{ela}
\end{eqnarray}
In a stack of layers the total elastic energy is composed of curvature
elastic energy Eq.\ref{ela} and the deformation energy due to smectic
dilations - compressions of the layers in the transverse direction,
characterised by a smectic compressibility modulus ${\cal B}$ \cite{prost}
\begin{equation}
H = \int H_n(K_{ik};{\bf n}) dn + {\textstyle{1\over 2}}{\cal B}
\int\!\!\!\int\left(\frac{\partial\zeta_n({\mbox{\boldmath $\rho $}} )}{\partial n}\right)^2 dn~d^2{\mbox{\boldmath $\rho $}}
\end{equation}
The formal limits of this expression are well known and will not be
discussed here. Introducing Fourier transform of the local
displacement in the directions ${\mbox{\boldmath $\rho $}}$ with wave vector $\bf Q$ we
obtain for the total smectic elastic energy the expression
\begin{eqnarray}
H &=& {\textstyle{1\over 2}}\!\sum_{\bf Q}\!\!\int\!\!dn~\left[ \left(
K_cQ^4 + a Q^2\left({\bf Q}{\bf n}(n)\right)^2 + b \left({\bf Q}{\bf
n}(n)\right)^4\right) \vert\zeta_n({\bf Q})\vert^2 + {\cal
B}\left(\frac{\partial\zeta_n(\bf Q )}{\partial n}\right)^2 \right] =
\nonumber\\ &=& {\textstyle{1\over 2}}\!\sum_{\bf
Q}\!\!\int\!\!dn~\zeta_n({\bf Q}){\cal H}(n,n';{\bf Q}) \zeta_n(-{\bf
Q})
\label{start}
\end{eqnarray}
where we have defined the operator ${\cal H}(n,n';{\bf Q})$ and used
the shorthand $\sum_{\bf Q} = \frac{S}{(2\pi )^2}\int d^2{\bf Q}$ with
$S$ being the area of the layer. The elastic energy expression
Eq. \ref{start} represents the final formalisation of our model. If
the intercalated polymers are chiral we have to add to Eq. \ref{start}
the standard cholesteric elastic energy which depends in the lowest
order only on the derivatives of the polymer director with respect to
the stack index \cite{prost}.

{\bf\sl Free energy of smectic fluctuations} We now proceed by
integrating out the smectic fluctuations from the free energy defined
through elastic Hamiltonian Eq.\ref{start} and thus obtaining an
effective intercalated polymer free energy that will depend only on
the director field of the polymers. We start by setting
\begin{equation}
\phi (n;{\bf Q}) = a Q^2\left({\bf Q}{\bf n}(n)\right)^2 + b \left({\bf Q}{\bf
n}(n)\right)^4
\end{equation}
and writing out explicitely the operator we introduced above Eq. \ref{start}
\begin{equation}
{\cal H}(n,n';{\bf Q}) = \left(-\beta{\cal B}\frac{\partial^2}{\partial
n^2} + \beta K_cQ^4 + \beta\phi (n;{\bf Q})\right)~\delta(n-n'),
\end{equation}
which allows us to express the free energy corresponding to smectic
elastic fluctuations as
\begin{eqnarray}
{\cal F}({\bf n}(n)) &=& -~kT~\ln{\Xi} = -~kT~\ln{\left( \Pi_{\bf
Q}\int\ddots\int {\cal D}\zeta_n({\bf Q})~exp{(-\beta~H)}\right)} =
\nonumber\\ &=& \frac{kT}{2}~\sum_{\bf Q} \ln{Det {\cal H}(n,n';{\bf
Q})}.
\label{free}
\end{eqnarray}
In order to get the part of the free energy that depends explicitely
on the polymer director field, it is convenient to resort to the
following representation of the free energy Eq.\ref{free}
\begin{eqnarray}
{\cal F}({\bf n}(n)) &=& \frac{kT}{2}~\sum_{\bf Q}~Tr~\ln{\left( {\cal
H}(n,n';{\bf Q})\right)} = \nonumber\\ &=& {\cal F}_0 +
\frac{1}{2}\sum_{\bf Q}Tr~\phi (n;{\bf Q})\int_{0}^{1} d\mu {\cal
G}_{\mu}(n,n';{\bf Q}), \nonumber\\ ~
\end{eqnarray}
where ${\cal F}_0$ is the part of the free energy that does not
contain ${\bf n}(n)$ explicitely. The Green function ${\cal
G}_{\mu}(n,n';{\bf Q})$ entering the above equation can be obtained
as \cite{chaikin} 
\begin{equation} 
\left( -\beta{\cal B}\frac{\partial^2}{\partial
n^2} + \beta K_cQ^4 + \mu\beta\phi (n;{\bf Q})\right)~{\cal G}_{\mu}(n,n';{\bf
Q}) = \delta(n-n').  
\end{equation} 
Introducing now ${\cal G}_{0}(n,n';{\bf Q}) = {\cal G}_{0}(\vert
n-n'\vert;{\bf Q}) = {\cal G}_{\mu = 0}(n,n';{\bf Q})$ in the form
\begin{equation}
{\cal G}_{0}(n-n';{\bf Q}) = {\textstyle{1\over 2}} \sqrt{\frac{\cal
B}{K_c}} Q^{-2}~\exp{\sqrt{\frac{K_c}{\cal B}}Q^2\vert n - n'\vert}
\end{equation}
we can expand ${\cal G}_{\mu}(n,n';{\bf Q})$ perturbatively up to the
first order in $\phi (n;{\bf Q})$ thus obtaining the free energy to
the second order in this quantity
\begin{equation}
\kern-50pt{\cal F}({\bf n}(n)) = {\cal F}_0 + \frac{1}{2}\sum_{\bf Q}~\int_{0}^{1}d\mu
\left[\int\!dn \phi (n;{\bf Q}){\cal G}_{0}(n,n;{\bf Q}) - \mu\beta
\int\!\!\int\!dndn' \phi (n;{\bf Q}){\cal G}_{0}^2(n,n';{\bf Q})\phi
(n';{\bf Q}) + \dots \right].
\label{expand}
\end{equation}
Since the zero order Green function is obviously of short range we
can expand the expression for the free energy Eq.\ref{expand} for a
slowly varying $\phi (n;{\bf Q})$ field in a standard fashion,
obtaining the following approximate form of the free energy
\begin{eqnarray}
{\cal F}({\bf n}(n)) &=& {\cal F}_0 + {\textstyle\frac{1}{2}}\sum_{\bf Q}\left[ {\cal
G}_{0}(0;{\bf Q})\int\!dn~\phi(n;{\bf Q}) \right. - \nonumber\\ &-&
{\textstyle{\beta\over 2}}\left(\int\! dt {\cal G}^2_{0}(t;{\bf Q})\right)
\int\!dn~\phi^2(n;{\bf Q}) + \nonumber\\ &+& \left. {\textstyle{\beta\over
4}} \left(\int\!dt~t^2{\cal G}^2_{0}(t;{\bf Q})\right)\int\!dn\left(
\frac{\partial\phi(n;{\bf Q})}{\partial n}\right)^2 \right].
\label{crap}
\end{eqnarray}
The summation over the Fourier space intends also integration over the
different directions of the ${\bf Q}$ vector. Denoting this
orientational integration with $\mathopen< \dots
\mathclose>_{{\mbox{\boldmath $\omega $}}}$ it is easy to see that $\mathopen<\phi(n;{\bf
Q})\mathclose>_{{\mbox{\boldmath $\omega $}}}$ as well as $\mathopen<\phi^2(n;{\bf
Q})\mathclose>_{{\mbox{\boldmath $\omega $}}}$ do not depend on the orientational angles
at all and are thus independent of the director field ${\bf n}(n)$. The
dependence on the director field remains only in the derivative terms,
{\sl i.e.} terms of the form $\left(\frac{\partial\phi(n;{\bf Q})}{\partial
n}\right)^2$. These terms contain
\begin{equation} 
\frac{d({\bf Q} {\bf n}(n))}{dn} = 
{\bf Q}\frac{d{\bf n}(n)}{dn} = {\bf Q}\dot{\bf n}(n) =
{\bf Q}\left( {\bf n}(n) \times {\mbox{\boldmath $\Omega $}} (n)\right)
\end{equation}
where ${\mbox{\boldmath $\Omega $}} (n)$ is the vector of the ``angular velocity'' of
rotation of the director from layer to layer. If the average normal to
the layers is in the $z$ direction than ${\mbox{\boldmath $\Omega $}} (n) =
(0,0,\Omega_z(n))$, also $\dot{\bf n}(n)$ is within each layer and in
direction prependicular to ${\bf n}(n)$ with magnitude $\vert
\dot{\bf n}(n)\vert = \Omega_z(n)$. As can be easily seen, the only
terms that survive the integration $\mathopen< \dots
\mathclose>_{{\mbox{\boldmath $\omega $}}}$ and still depend on the director field are
those depending quadratically on $\vert \dot{\bf n}(n)\vert$. From the
free energy Eq.\ref{crap} these terms can be obtained in the form
\begin{equation}
{\cal F}({\bf n}(n)) = {\cal F}_0(a,b) + {\frac{\beta}{8}} \frac{S}{4\pi}
\int_0^{\infty}\!\!Q^9 dQ \left(\int\!dt~t^2{\cal G}^2_{0}(t;{\bf
Q})\right) \left( (a+b)^2 + \left(\frac{b}{2}\right)^2\right) \int
\vert\dot{\bf n}(n)\vert^2 dn,
\end{equation}
where ${\cal F}_0(a,b)$ is the part of the free energy that after the $\mathopen< \dots
\mathclose>_{{\mbox{\boldmath $\omega $}}}$ integration does not
depend explicitely on the director any more. Evaluating the last integral
over $Q$ and taking into account the fact that the minimal value of
$Q$ is set by the domain size, assumed to be a square of side $l$,
while the maximal value is set by the molecular dimension $a$, we
obtain finally
\begin{eqnarray}
{\cal F}({\bf n}(n)) &=& {\cal F}_0(a,b) + \frac{\beta~S}{512 \pi}
\left(\frac{\cal B}{K_c}\right)^{5/2} \left( (a+b)^2 +
\left(\frac{b}{2}\right)^2\right) \ln{\frac{l}{a}} \int\vert\dot{\bf n}(n)
\vert^2 dn = \nonumber\\
&=& {\cal F}_0(a,b) + \frac{K_{st}}{2} \int\vert\dot{\bf n}(n)\vert^2
dn
\label{konec}
\end{eqnarray}
Clearly for this geometry $K_{st}$ is the additional twist elastic
constant of intercalated polymers, stemming from the smectic
interactions between layers of orientationally ordered polymer
molecules. Thus we see that in an intercalated smectic the smectic
interactions tend to renormalize the twist elastic modulus of
intercalated oriented polymers to
\begin{equation}
K_2 \longrightarrow K_2 + \frac{\beta}{256\pi} \left(\frac{\cal
B}{K_c}\right)^{5/2} \left( (a+b)^2 +
\left(\frac{b}{2}\right)^2\right)  \ln{\frac{l}{a}},
\label{twist}
\end{equation}
where $K_2$ is the ``bare'' polymer twist modulus, {\sl i.e.} the
twist elastic modulus of the polymer subphase if the smectic
interactions are not taken into account. This renormalization is of
purely fluctuational origin. If the polymers are in addition chiral,
the renormalization of the twist modulus obviously leads to unwinding
of the cholesteric pitch of the intercalated polymers to a new
equilibrium value of
\begin{equation}
P \longrightarrow P \left( 1 + \frac{\beta}{256\pi K_2} \left(\frac{\cal
B}{K_c}\right)^{5/2} \left( (a+b)^2 +
\left(\frac{b}{2}\right)^2\right) \ln{\frac{l}{a}}\right).
\end{equation}

{\bf\sl Discussion} The mean - field model introduced above led in a
very straightforward way to a renormalization of the twist modulus of
intercalated polymers. The effect itself is a very intuitive one. If
there exists a deformational ``easy axis'' within each smectic layer,
which according to Eq.\ref{start} represents a deformational wave
whose direction is perpendicular to ${\bf n}(n)$ within a single
layer, the smectic compressibility term would tend to twist the
directors of the neighboring layers towards a colinear position. This
would introduce a coupling term $\left({\bf n}(n) - {\bf
n}(n+1)\right)^2$ in the free energy of two neighboring layers. The
continuum version of this effect would lead exactly to Eq.\ref{konec}.

The coupling constants $a$ and $b$, {\sl c.f.} Eq.\ref{ela}, between
the orientational ordering of intercalated polymers and effective
elastic properties of the layers depend in general on the
orientational order parameter of the polymer chain $\cal S$, the
elastic modulus of the chains defined as $kT {\cal L}_p$ where ${\cal
L}_p$ is the persistence length and the polymer surface density
$\rho$. The orientational order parameter of the polymer chains is
defined through the 2-D orientational tensor $\sigma_{ik}$ of the
intercalated chains as $\sigma_{\alpha} = \frac{\rho}{2}( 1 \pm {\cal
S})$, where $\alpha$ is the index of the two eigenvalues
\cite{rudi}. For $\cal S$ close to $1$, {\sl i.e.} close to complete
orientational order where all the chains point in the same direction,
the scaling form for $a$ and $b$ should be
\begin{equation}
a,b \sim kT~{\cal S} \frac{{\cal L}_p}{\ell_\perp},
\label{ab}
\end{equation}
where $\ell_\perp$ is the separation between the chains perpendicular
to their long axis. This is the form appropriate for our assumption of
complete ordering of chains within each smectic layer.

We can now asses the magnitude of the contribution of smectic modes to
the twist elastic modulus of the smectic layers. Assuming the scaling
Eq. \ref{ab} for constants $a$ and $b$ we obtain
\begin{equation}
K_{st} \sim (kT)^{-1} \left(\frac{\cal B}{K_c}\right)^{5/2} \left( kT~{\cal
S} \frac{{\cal L}_p}{\ell_\perp}\right)^2
\end{equation}
times an unknown numerical constant persumably on the order of
$10$. In the units used in Eq. \ref{start} $\frac{\cal B}{K_c}$ is
dimensionless. Since the dominant interactions determining $\cal B$
are electrostatic attractions between DNA and cationic lipid
headgroups the smectic modulus has to be quite large while
$\frac{{\cal L}_p}{\ell_\perp}$ is on the order of $10$, the
modification of the twist modulus implied by Eq. \ref{twist} thus has
to be enormous. If the intercalated polymers are chiral it would thus
come as no surprise if the effective cholesteric pitch surviving in
this system would be orders of magnitude larger than in the pure
polymer system \cite{joachim}.

We have not included the possible direct modification of the chiral
interactions by smectic fluctuations in this analysis, an effect that
would act in the direction opposite to the smectic fluctuation
renormalization of the polymer twist modulus. As this effect would
tend to make the effective pitch smaller, which apparently has never
been observed in this system, we assume that it is small.

In conclusion we have shown how the smectic degrees of freedom couple
to orientational modes of orientationally ordered polymers
intercalated between smectic layers leading to fluctuation
renormalization of the polymer twist modulus. We derived a substantial
renormalization of the effective cholesteric pitch in such a system,
if the polymers themselves are chiral. We propose this as the primary
reason why no cholesteric structures have been observed in the DNA-CL
system.

{\bf\sl Acknowledgement} One of the authors, R.P., would like to
acknowledge numerous discussions with Helmut Strey of DCRT/NIH on different
features of the intercalated smectic phases.

{\bf \sl Figures} Fig.1. A schematic representation of a part of a
polymer intercalated smectic system. The smectic layers are not shown
explicitely. The direction of the intercalated polymers changes from
layer to layer. It has been arbitrarily assumed to change by
$\frac{\pi}{4}$ between two neighboring layers. The smectic
interactions tend to orient the neighboring polymer layers in a
parallel direction.



\begin{thebibliography}{99} 
\bibitem{dnacl1} D.D. Lasi\v c, H.H. Strey, M.C.A. Stuart, R. Podgornik and
P.M. Frederik, {\sl J. Amer. Chem. Soc.} 119, 832-833 (1997).
\bibitem{dnacl2} J.O. R\" adler, I. Koltover, T. Salditt, C.R. Safinya, {\sl Science} 
275, 810-814 (1997).  
\bibitem{adrian1} V.A. Parsegian and R.P. Rand,
in {\sl Structure and Dynamics of Membranes}, Handbook of Biological
Physics, Vol. 1, Eds. R. Lipowsky and E. Sackmann, (Elsevier) 643 -
690 (1995).  
\bibitem{prl} H.H. Strey, D.C. Rau, V.A. Parsegian and R. Podgornik, 
{\sl Phys.Rev.Lett.} 78 895-898 (1997)
\bibitem{pnas} R. Podgornik, H.H. Strey, K. Gawrisch, D.C. Rau, A. Rupprecht
and V.A. Parsegian, {\sl Proc. Natl. Acad. Sci.} 93, 4261-4266 (1996).
\bibitem{joachim} J.O. R\" adler, personal communication (1997).
\bibitem{jie} J. Yang, J. Mou and Z. Shao, {\sl FEBS Letters} 301,
173-175 (1995).  
\bibitem{david} F. David, E. Guitter and L. Peliti,
{\sl J. Physique} 48, 2059-2066 (1987).  
\bibitem{prost} P.G. de Gennes and J. Prost, {\sl The Physics of Liquid Crystals}, 
(Oxford) (1993).  
\bibitem{chaikin} P.M. Chaikin and T.C. Lubensky, {\sl Principles of Condensed Matter}, (Cambridge) (1995).
\bibitem{rudi} R. Podgornik, {\sl Phys.Rev.~E} 54, 5268-5277 (1996). 
\end{thebibliography}
\end{document}